\documentclass[9pt]{book}

\usepackage[dvips]{graphicx,color}
\usepackage{makeidx,universe}

\makeauthorindex

\BookTitle{The proceedings of the Physics of 
           Accreting Compact Binaries}

\CopyRight{\copyright 2010 by Universal Academy Press, Inc.}

\begin{document} 

\pagenumbering{arabic}

\chapter{Eclipse Mapping the Flickering Sources 
 In The Dwarf Nova HT Cassiopeia}

\author{\raggedright \baselineskip=10pt%
{\bf R.\ Baptista,$^{1}$, B.\ Borges,$^{2}$,
V.\ Kolokotronis,$^{3}$, O.\ Giannakis,$^{3}$
and C.\ J.\ Papadimitriou$^{3}$}\\ 
{\small \it %
(1) Departamento de F\'{\i}sica, UFSC, Campus Trindade, 
    Florian\'opolis, 88040-900, Brazil \\
(2) Faculdade de Ci\^encias Exatas e Tecnologia, UFGD, Dourados,
    79825-070, Brazil  \\
(3) Institute of Astronomy and Astrophysics, NOA, Thissio 11810,
    Athens, Greece
}
}

\AuthorContents{R.\ Baptista, B.\ Borges, V. Kolokotronis, 
 O.\ Giannakis, and C.\ J.\ Papadimitriou} 

\AuthorIndex{Baptista}{R.}
\AuthorIndex{Borges}{B.}
\AuthorIndex{Kolokotronis}{V.}
\AuthorIndex{Giannakis}{O.}
\AuthorIndex{Papadimitriou}{C.J.}

     \baselineskip=10pt
     \parindent=10pt

\section*{Abstract} 

We report results of the eclipse mapping analysis of an
ensemble of light curves of HT Cas.
The fast response of the white dwarf (WD) to the increase in mass
transfer rate, the expansion rate of the accretion disc at the
same time, and the relative amplitude of the high-frequency
flickering indicate that the quiescent disc of HT~Has has high
viscosity, $\alpha\simeq 0.3-0.7$. This is in marked disagreement
with the disc-instability model and implies that the outbursts of
HT~Cas are caused by bursts of enhanced mass-transfer rate from
its donor star.

\section{Introduction} 

Flickering is the intrinsic brightness fluctuation of 0.01-1~mag
on timescales of seconds to dozens of minutes that is seen in
light curves of T~Tau stars, mass-exchanging binaries and active
galactic nuclei. It is considered a basic signature of accretion, and
may be used to probe the anomalous viscosity that powers accretion
discs. Earlier studies \cite{refbapht.4} led to the suggestion
that there are mainly two sources of flickering in Cataclysmic
Variables: the stream-disc impact region at disc rim and a turbulent
inner disc region in the vicinity of the WD. More recent studies
found that low- and high-frequency flickering may arise at different
locations and be related to distinct causes \cite{refbapht.1}.

HT~Cas is a well-studied 106~min period eclipsing SU~UMa type dwarf
nova with strong flickering activity, probably arising from close
to its WD \cite{refbapht.4}. Aside of its infrequent outbursts
($V\sim 13$~mag), it shows transitions from a bright ($V\simeq
16.4$~mag) to a faint state ($V\simeq 17.7$~mag) on timescales of
days to months. This behaviour is reminiscent to that of the
VY~Scl novalike stars, and is similarly interpreted in terms of a
significant change in mass-transfer rate \cite{refbapht.7}.

\section{Observations and Data Analysis} 

Time-series of white-light photometry of HT Cas were collected
between 2007 and 2009 with the 1.2~m telescope at the Astronomical
Station Kryoneri, in Greece. The data comprise 63 eclipse light
curves and frame a 2\,d long transition from faint to bright state
in 2007 January. The light curves of HT~Cas and of a comparison star
of similar brightness are shown in Fig.\,1. The scatter with respect
to the mean level is significantly larger in HT~Cas and is caused by
a combination of flickering and long-term brightness changes. The
scatter is larger for increasing brightness state, indicating that
the flickering amplitude varies with mass transfer rate.
There is a clear reduction of scatter during the eclipse of the WD.
%
\begin{figure}[t] 
 \begin{center}
  \includegraphics[width=0.73\textwidth,height=\textwidth,angle=-90]
   {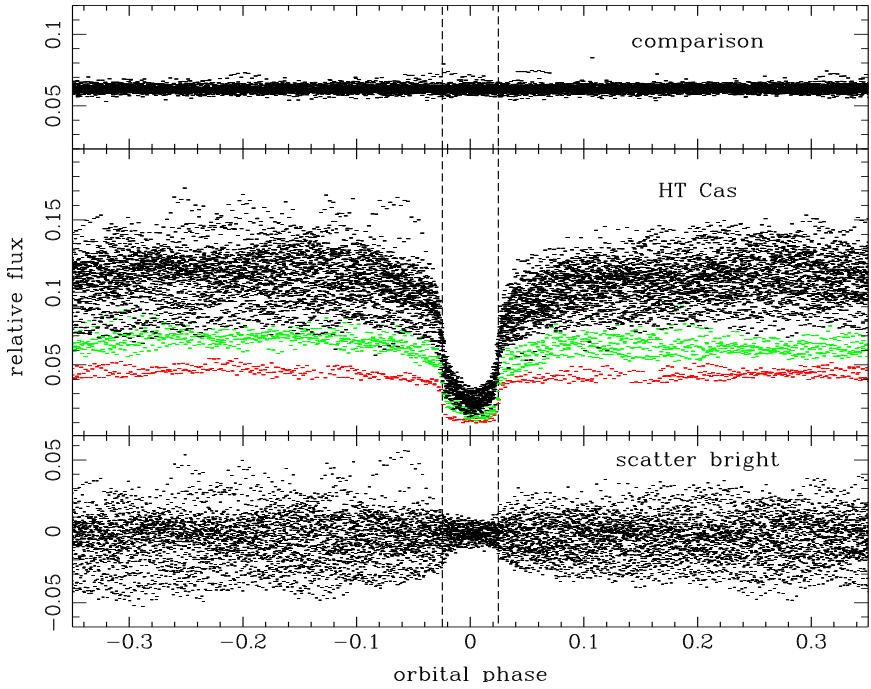}
  \caption{Ensemble of light curves of HT~Cas. Top: the light
  curves of a comparison star of similar brightness; the scatter is
  caused by Poisson noise. Middle: light curves of the bright
  (black), faint (red) and intermediate transition (green) states.
  Bottom: light curves of the bright state after subtraction of 
  the average orbital curve. Vertical dashed lines mark the
  ingress/egress phases of the WD.}
 \end{center}
\end{figure}

We applied the 'single' and 'emsemble' methods to the light curves of
the bright and intermediate states to derive the orbital dependency
of its steady-light, long-term changes, and low- and high-frequency
flickering components \cite{refbapht.1}. In order to derive maps of
the accretion disc only, the contribution of the WD was removed
from the steady-light light curves before applying the eclipse
mapping method.

\section{Results}

The rise from the faint to the intermediate state takes 1\,d, and
from that to the bright state another 1\,d. Eclipse maps show
enhanced emission along the stream trajectory in the intermediate
and bright states (Fig.\,2), indicating that the rise in brightness
is caused by an increase in mass transfer rate. In response to that,
the disc increases in brightness by a factor 3, and the WD at disc
center brightens by a factor 2 -- probably as a consequence of
accretional heating \cite{refbapht.6}.
The newly added disc gas reaches the WD at disc center quickly
after mass transfer recovery ($\sim 1$\,d), implying a disc
viscosity parameter $\alpha_{\rm quies}\simeq 0.5$. Furthermore, 
the disc expands (at $v\simeq +0.4\, km\,s^{-1}$) in response to
the increased mass transfer rate, also implying 
$\alpha_{\rm quies}\simeq 0.3-0.5$. Both results point to a highly
viscous quiescent disc in HT~Cas, in marked contrast to the 
expectations of the disc-instability model ($\alpha_{\rm quies}\sim
10^{-2}$) \cite{refbapht.8}.
%
\begin{figure}[t] 
 \begin{center}
  \includegraphics[height=0.95\textwidth,angle=-90]{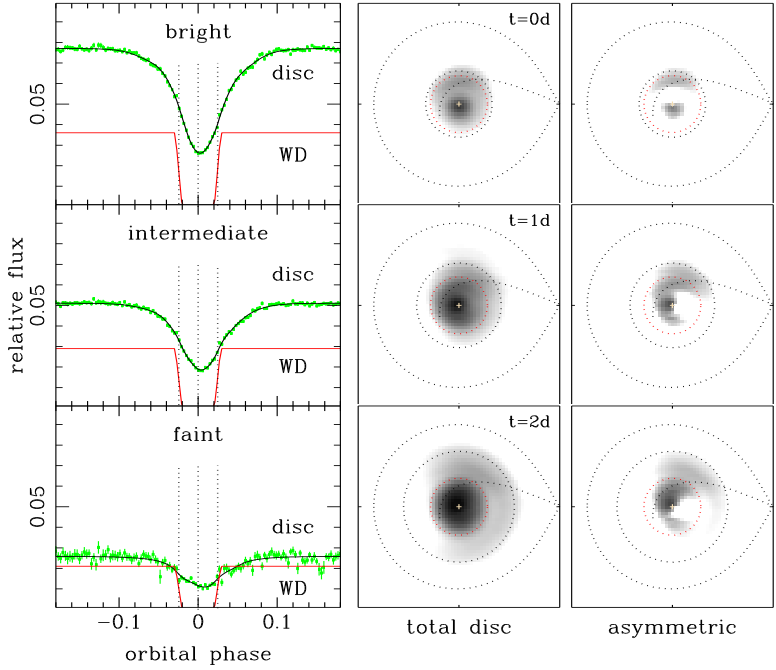}
  \caption{Left: WD-subtracted steady-light data for the faint, 
   intermediate and bright states (green dots) and model (black
   lines) light curves. The extracted light curve of the WD is shown
   as a red line in each case. Vertical dotted lines mark 
   ingress/egress phases of WD and mid-eclipse. Center: eclipse maps
   in a logarithmic grayscale; brighter regions are indicated in
   black, fainter regions in white. Black dotted lines show the
   primary Roche lobe, the gas stream trajectory, and the inferred
   disc radius in each state. A red dotted circle depicts the
   circularization radius. Right: asymmetric component of the
   eclipse maps in the center panels.}
 \end{center}
\end{figure}

While the increase in mass transfer rate leads to clear gas stream
overflow, there is no corresponding increase in the orbital hump
modulation (to signal enhanced gas inflow in the bright spot at
disc rim). A high viscosity disc has relatively low density and is
easily penetrated by an enhanced gas stream flow \cite{refbapht.3}.
In these cases, the common idea that pronounced orbital hump
modulation would be expected in response to an increase in mass
transfer rate \cite{refbapht.8} is misleading.

Fig.\,3 shows 'ensemble' and 'single' flickering maps and the
corresponding radial dependency of their relative amplitude. 
\begin{figure}[t] 
 \begin{center}
  \includegraphics[height=0.9\textwidth,angle=-90]{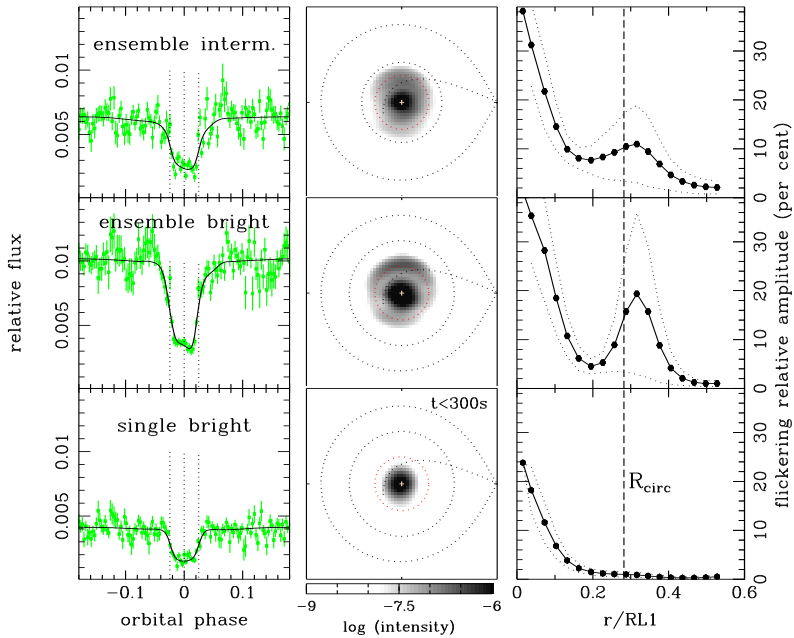}
  \caption{Left: ensemble and single data (green dots) and model
  light curves. Center: eclipse maps in a logarithmic grayscale.
  The notation is similar to that of Fig.3. The scalebar
  depicts the log(intensity) scale. Right: Relative amplitude of the
  flickering as a function of radius (in units of RL1, the distance
  from WD to the inner Lagrangian point), derived by dividing each
  flickering map by the steady-light map of the corresponding
  brightness state. Dotted lines show the 1-$\sigma$ limit on the
  amplitude, and a vertical dashed line indicate the circularization
  radius.}
 \end{center}
\end{figure}
High-frequency ('single') flickering is concentrated in the
innermost disc regions, whereas low-frequency ('ensemble' minus
'single') flickering is distributed over a larger region, with
enhanced emission from near the circularization radius. The WD does
not flicker. If one divides the flickering maps by steady-light maps
including the WD the relative flickering amplitude goes to zero at
disc center. On the other hand, dividing the flickering maps by the
corresponding disc-only steady-light maps reveals that the flickering
amplitude increases sharply towards disc center. The ensemble
flickering amplitude reaches $\sim 40$\% near disc center. The
low-frequency flickering at $R\sim R_{\rm circ}$ may be related to
the mass-transfer process (unsteady gas inflow or turbulence/shocks
generated at the impact of infalling matter with disc gas
\cite{refbapht.4}).

Assuming that the inner disc flickering is caused by fluctuations in
energy dissipation rate induced by MHD turbulence \cite{refbapht.5},
its relative amplitude yields a direct measurement of the disc
viscosity parameter and its radial dependency. We find $\alpha(r)
\simeq 0.7\,[r/(0.1\,R_{L1})]^{-2}\, [52\, H/r]^{-1}$, in agreement
with our previous estimates.

\section{Conclusions}

The fast response of the WD to the increase in mass transfer
rate and the expansion rate of the disc from the faint to the bright
quiescent states imply a highly viscous disc in HT Cas, 
$\alpha_{\rm quies}\simeq 0.3-0.5$, in disagreement with predictions
of the disc-instability model.
High-frequency flickering arises from the inner disc regions and
seems connected to turbulence in the disc itself. The low-frequency
flickering shows a more extended distribution, with a ring of
emission near the circularization radius. Both components rise
sharply towards disc center.
If inner disc flickering is caused by MHD turbulence, the disc
viscosity decreases with radius as $\alpha(r) \propto r^{-2}$.


\end{document}